**Three-dimensional alteration of neurites in schizophrenia**

Ryuta Mizutani[1]*, Rino Saiga[1], Akihisa Takeuchi[2], Kentaro Uesugi[2], Yasuko Terada[2], Yoshio Suzuki[3], Vincent De Andrade[4], Francesco De Carlo[4], Susumu Takekoshi[5], Chie Inomoto[5], Naoya Nakamura[5], Itaru Kushima[6], Shuji Iritani[6], Norio Ozaki[6], Soichiro Ide[7], Kazutaka Ikeda[7], Kenichi Oshima[7], Masanari Itokawa[7], and Makoto Arai[7]

[1]Department of Applied Biochemistry, Tokai University, Hiratsuka, Kanagawa 259-1292, Japan.
[2]Japan Synchrotron Radiation Research Institute (JASRI/SPring-8), Sayo, Hyogo 679-5198, Japan.
[3]Graduate School of Frontier Sciences, University of Tokyo, Kashiwa, Chiba 277-8561, Japan
[4]Advanced Photon Source, Argonne National Laboratory, Lemont, IL 60439, USA.
[5]Tokai University School of Medicine, Isehara, Kanagawa 259-1193, Japan.
[6]Graduate School of Medicine, Nagoya University, Nagoya, Aichi 466-8550, Japan.
[7]Tokyo Metropolitan Institute of Medical Science, Setagaya, Tokyo 156-8506, Japan.

*Correspondence should be addressed to:
Department of Applied Biochemistry, Tokai University, Hiratsuka, Kanagawa 259-1292, Japan
Phone: +81-463-58-1211; Fax: +81-463-50-2426
E-mail: mizutanilaboratory@gmail.com

**Abstract**

Psychiatric symptoms of schizophrenia suggest alteration of cerebral neurons. However, the physical basis of the schizophrenia symptoms has not been delineated at the cellular level. Here we report nanometer-scale three-dimensional analysis of brain tissues of schizophrenia and control cases. Structures of cerebral tissues of the anterior cingulate cortex were visualized with synchrotron radiation nanotomography. Tissue constituents visualized in the three-dimensional images were traced to build Cartesian coordinate models of tissue constituents, such as neurons and blood vessels. The obtained Cartesian coordinates were used for calculating curvature and torsion of neurites in order to analyze their geometry. Results of the geometric analyses indicated that the curvature of neurites is significantly different between schizophrenia and control cases. The mean curvature of distal neurites of the schizophrenia cases was approximately 1.5 times higher than that of the controls. The schizophrenia case with the highest neurite curvature carried a frame shift mutation in the GLO1 gene, suggesting that oxidative stress due to the GLO1 mutation caused the structural alteration of the neurites. The differences in the neurite curvature result in differences in the spatial trajectory and hence alter neuronal circuits. It has been shown that the anterior cingulate cortex analyzed in this study has emotional and cognitive functions. We suggest that the structural alteration of neurons in the schizophrenia cases should reflect psychiatric symptoms of schizophrenia.

**Introduction**

Schizophrenia is a chronic mental disorder that affects approximately 1% of the population.[1] Clinical symptoms of schizophrenia include hallucinations, delusions, emotional disorders, and cognitive dysfunction. The development of these symptoms suggests alterations in the connectivity between cerebral neurons. It has been reported that dendritic spines of neurons are significantly decreased in the external pyramidal layer of the cerebral cortex of schizophrenic brains.[2-4] Since dendritic spines form the majority of excitatory synapses, the loss of spines can directly impair neuronal connectivity. The reduced neuropil hypothesis[5] posits that reductions in neuron size and arborization are the explanation for the reduced brain volume observed in schizophrenia.[6-9] The reductions in neuron size and arborization can perturb the neuronal structures, resulting in changes to the neuronal circuits. However, studies of brain tissues of schizophrenia patients have mainly been performed using two-dimensional images of tissue sections, whereas the neurons themselves are three-dimensional in nature.

It has been reported that three-dimensional structures of brain tissues can be analyzed with electron microscopy by reconstructing them from serially sectioned images.[10-12] Since soft tissues are deformed by sectioning, the deformations are artificially corrected in the three-dimensional reconstruction.[13,14] Therefore, three-dimensional image reconstructed from serial sections does not exactly reproduce the three-dimensional structure of the tissue. Another method to visualize the three-dimensional structure of biological tissue is confocal light microscopy. However, light microscopy cannot visualize structures behind opaque objects. Its resolution is three-dimensionally anisotropic and depends on the direction of the optical axis.[15] Although neuronal structures have been deposited in the NeuroMorpho.Org database[16] including those of human cerebral cortex,[17] three-dimensional coordinates estimated from light microscopy images show irregular displacements especially along the optical axis and are not applicable to geometric analyses. Thus, resolution anisotropy, tissue opacity, and sectioning deformation can degrade the three-dimensional features that may be relevant to schizophrenia.

In this study, we analyzed brain tissue structures of schizophrenia patients and control cases with synchrotron radiation nanotomography.[18,19] X-ray microtomography and nanotomography[20-25] can visualize three-dimensional opaque objects with nearly isotropic resolution.[26] Its reconstruction process does not involve any deformation correction, and hence, the obtained image reproduces the actual three-dimensional structures. Although the low x-ray contrast of the brain tissue itself initially limited visualizations to those of large-scale structures,[27] the use of high-Z element staining has since allowed neuronal networks in brain tissue to be observed.[28,29] Tissue constituents visualized in this study were traced to build three-dimensional Cartesian coordinate models of tissue structures, such as neurons and blood vessels. The three-dimensional tissue structures were reproduced as Cartesian coordinates through this process. The resultant neuronal coordinates can be used for analyzing the geometry of neurons in schizophrenia and control cases.

Cerebral tissues analyzed in this study are those of the anterior cingulate cortex (Brodmann area 24). It has been shown that this brain area has emotional and cognitive function.[30,31] It has also been reported that the anterior cingulate cortex is related to schizophrenia,[32,33] attention-deficit/hyperactivity disorder,[34] and obsessive-compulsive disorder.[35] Therefore, the structural character of neurons of the anterior cingulate cortex can affect the mental capabilities or the psychiatric symptoms. In this study, we examined differences of neuronal structures between schizophrenia and control cases and also between control individuals.

**Materials and Methods**
**Cerebral tissue samples**

All post-mortem human cerebral tissues were collected with informed consent from the legal next of kin using protocols approved by the clinical study reviewing board of Tokai University School of Medicine (application no. 07R-018) and the ethics committee of Tokyo Metropolitan Institute of Medical Science (approval no. 17-18). This study was conducted under the approval of the ethics committee for the human subject study of Tokai University (approval nos. 11060, 11114, 12114, 13105, 14128, 15129, 16157, and 18012). Schizophrenia patients S1–S4 (Supplementary Table S1) were diagnosed according to the DSM-IV codes with the consensus of at least two experienced psychiatrists. Control patients (Supplementary Table S1) had been hospitalized because of a traffic injury (N1) or non-psychiatric lethal diseases (N2–N4) and were not psychiatrically evaluated. Since the cause of death of the N1 case was damage to the heart, histological changes in brain tissue specific to the injury can be excluded. The number of cases was determined in consideration of the available beamtime at the synchrotron radiation facilities. The control cases were selected so as to match the gender and age of the schizophrenia cases. Cases in which hemorrhage, infarction, or neoplasm were observed in the histological assessment of cerebral tissues were excluded. No previous records of schizophrenia were found for the control cases. The cerebral tissues of the anterior cingulate cortex (Brodmann area 24) were collected from the left hemispheres of the biopsied brains and subjected to Golgi impregnation, as described previously,[29] in order to visualize neurons in x-ray images.

The Golgi-stained tissues were first soaked in neat ethanol, then in *n*-butyl-glycidyl ether, and finally in Petropoxy 154 (Burnham Petrographics, USA) epoxy resin, as described previously.[36] The resin-soaked tissues were cut into rod shapes with approximate widths of 0.5 mm and lengths of 3–5 mm under a stereomicroscope and then transferred to borosilicate glass capillaries (W. Müller, Germany) filled with resin. The capillary diameter was approximately 0.8 mm. The capillaries were incubated at 90°C for 70–90 hours for curing the resin.

**Nanotomography**

The distal end of the capillary sample was sleeved with a brass tube using epoxy glue and secured with a setscrew to a brass or invar adapter specially designed for nanotomography. The mounted samples were placed in the experiment hutch as soon as possible in order to equilibrate their temperature with that of the sample stage. Nanotomography experiments using Fresnel zone plate optics were performed at the BL37XU[18] and BL47XU[37] beamlines of the SPring-8 synchrotron radiation facility and at the 32-ID beamline[19] of the Advanced Photon Source (APS) of the Argonne National Laboratory. Since the nanotomography experiments were performed in the manner of local computed tomography (CT), only the region of interest within the viewing field (64–122 μm diameter; Supplementary Table S2) was visualized.

In the experiments at the SPring-8 beamlines, the tissue samples were mounted on a slide-guide rotation stage specially built for nanotomography (SPU-1A, Kohzu Precision, Japan). Transmission images were recorded with a CMOS-based imaging detector (ORCA-Flash4.0, Hamamatsu Photonics, Japan) using monochromatic radiation at 8 keV. Examples of raw images and their reconstructed slices are shown in Supplementary Figure S1. Octagonal sector condenser zone plates[37] were used as beam condensers. The photon flux at the sample position of the BL37XU optics of the 2017.4 setup was estimated to be $1.0 \times 10^{14}$ photons/mm$^2$/s using $Al_2O_3$:C dosimeters (Nagase-Landauer, Japan). Since the specific gravity of Petropoxy 154 is 1.18 and its attenuation coefficient was estimated to be 6.1 cm$^{-1}$ at 8 keV, this x-ray flux corresponds to an absorbed radiation dose by the resin of $7.0 \times 10^4$ Gy/s. In the nanotomography experiments at the APS beamline, the tissue samples were mounted on an air-bearing rotation stage (UPR-160AIR, PI miCos, Germany) during the 2016.6 beamtime, or on a motorized model 4R Block-Head air bearing spindle (Professional Instruments Company, USA) during the 2017.10 beamtime. Zernike phase-contrast images were recorded with a CMOS-based imaging detector (GS3-U3-51S5M-C, FLIR, USA) using monochromatic radiation at 8 keV. A polygonal beam shaping condenser and compound refractive lenses were used as beam condensers.[19] Spatial resolutions were estimated using three-dimensional square-wave patterns[38] or from the Fourier domain plot.[39] The experimental conditions are summarized in Supplementary Table S2. The data collection procedure is described in the Supplementary Materials and Methods.

While absorption contrast was sufficient for visualization of the neurons in the SPring-8 BL37XU experiments, the Zernike phase-contrast method was used in order to enhance the sample contrast in the experiments at the BL47XU beamline of SPring-8 and at the 32-ID beamline of APS. Although a number of contrast-enhancing methods have been reported,[20,27,40] methodological prerequisites, such as the strict requirement of x-ray coherence in the holography, pose limitations in real applications. In this study, approximately $4 \times 10^5$ images of human brain tissues were collected in order to examine the statistical significance of their structural differences. Hence, practical aspects including the time efficiency and the methodological tolerance against the beam fluctuation should be taken into account. For these reasons, we used the Zernike method to enhance the sample contrast.

**Tomographic reconstruction**

Tomographic slices perpendicular to the sample rotation axis were reconstructed with the convolution-back-projection method using the RecView software.[36] The reconstruction calculation was performed by RS. Image pixels of APS datasets were averaged by 2 × 2 binning prior to the tomographic reconstruction, since the original pixel size of 26 nm was approximately half the pixel sizes of the SPring-8 experiments (40.2, 48.3 and 59.6 nm, respectively; Supplementary Table S2) and the binned pixel size of 52 nm was sufficiently fine compared to the spatial resolution of the APS datasets (300 nm). Since the reconstruction of the SPring-8 datasets was performed without binning, the reconstructed voxel size was the same as the original pixel size. Multiple image sets taken by shifting the sample were aligned and stacked to obtain the entire three-dimensional image. The entire reconstructed volume of each dataset was subjected to further analysis.

**Data blinding**

Nanotomography datasets were analyzed with the role allotment of data management to RS and data analysis to RM. The data manager reconstructed tomographic slices of all datasets and coded the dataset name. The data manager managed the information regarding the case assignment of each dataset but had no access to the analysis results except for an indicator of the analysis amount. The data analyst built Cartesian coordinate models from the datasets but had no access to the case information. The whole model building procedure was performed by RM in order to keep the modeling quality constant. The detailed protocol is described in the Supplementary Materials and Methods.

**Cartesian coordinate models and geometric analysis**

Cartesian coordinate models were built in four steps: (1) manual assignment of large structural constituents such as somata and blood capillary vessels, (2) automatic tracing to build a computer generated model and its subsequent refinement, (3), examination of the entire three-dimensional image and manual intervention to modify the working model, and (4) final structural refinement. The geometric analysis was performed in three steps: (1) cell typing, (2) structure annotation, and (3) geometric parameter calculation. Detailed protocols are described in Supplementary Materials and Methods.

**Code availability**

The RecView software[36] that was used for the tomographic reconstruction is available from https://mizutanilab.github.io under the BSD 2-Clause License. The model building and geometric analysis procedures were implemented in the MCTrace software[41] available from https://mizutanilab.github.io under the BSD 2-Clause License.

**Results**

**Three-dimensional tissue structure**

The cerebral cortex tissues analyzed in this study were taken from the left hemispheres of autopsied brains of four schizophrenia cases S1–S4 (age: mean 65 ± standard deviation 6 yr; 2 males and 2 females) and four age/gender-matched control cases N1–N4 (64 ± 5 yr; 2 males and 2 females). The N1 tissue may have suffered from a prolonged post-mortem time (85 h; Supplementary Table S1) compared with the other cases. Three-dimensional structures of these cerebral tissues were visualized with nanotomography[18,19] (Fig. 1a and Supplementary Figure S2) at spatial resolutions of 180–300 nm (Supplementary Table S2). Tissue constituents, such as neurites and blood vessels visualized in the 55 obtained datasets, were traced to build three-dimensional Cartesian coordinate models of the tissues (Fig. 1b–d and Supplementary Figs. S3–S4). An example of a spiny dendrite superposed on a three-dimensional map is shown in Fig. 1e. Representative images and models of all datasets are shown in Supplementary Figs. S2–S4. The statistics of the datasets and models are summarized in Supplementary Table S3. Coordinate files are available from https://mizutanilab.github.io (RRID:SCR_016529).

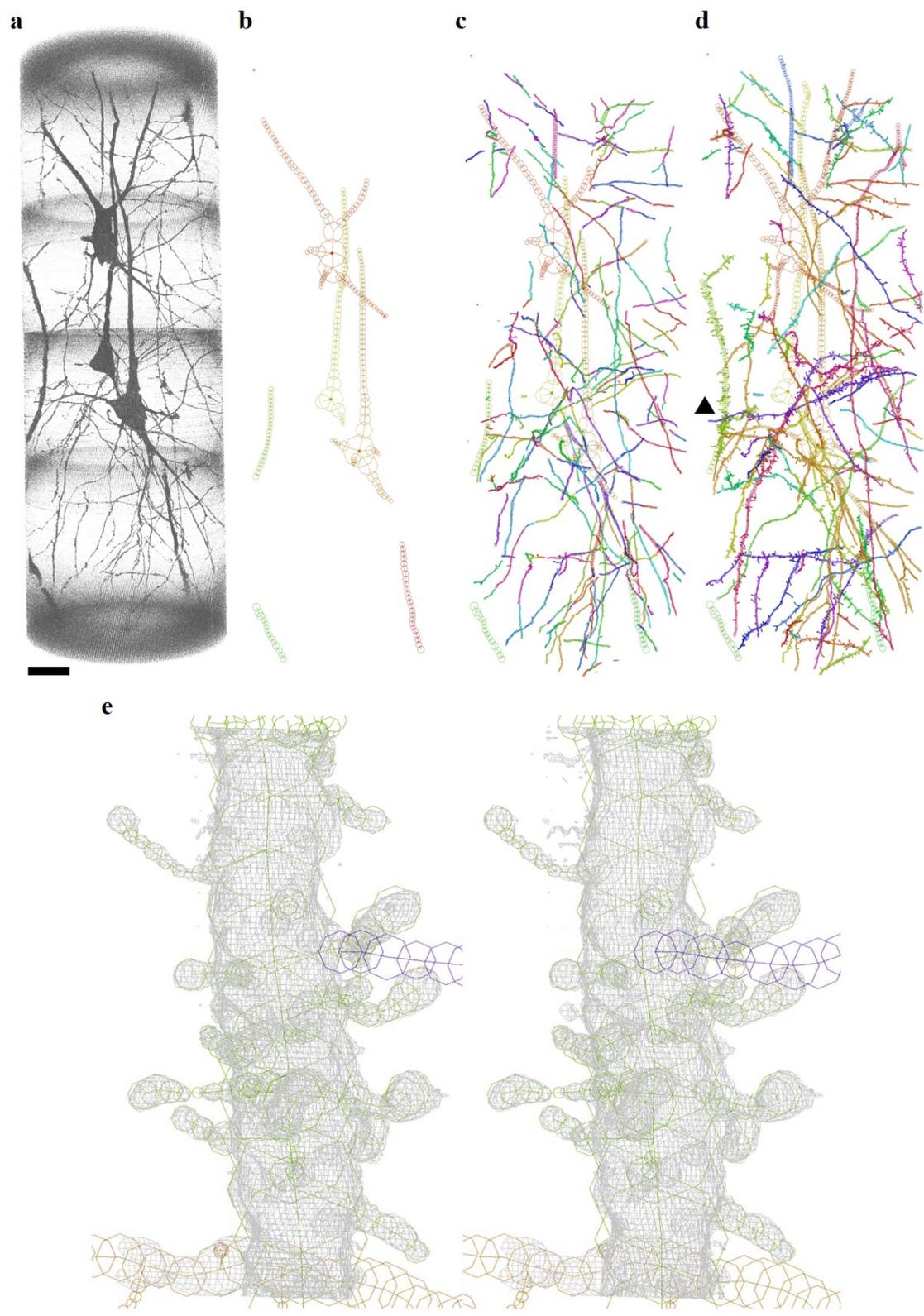

**Figure 1.** Three-dimensional visualization of cerebral cortex neurons and their models represented with Cartesian coordinates. The pial surface is toward the top. The three-dimensional image was rendered with VGStudio (Volume Graphics, Germany).

The models were drawn using MCTrace.[41] Structural constituents of the model are color-coded. Nodes composing each constituent are indicated with circles. Dots indicate soma nodes. (**a**) Rendering of dataset N2C of the control N2 tissue. Voxel values of 500–1600 were rendered with the scatter HQ algorithm. Scale bar: 20 μm. (**b**) Initial model. Structures of somata and thick neurites were built manually in order to mask them in the subsequent automatic model generation. (**c**) Automatically generated model of the tissue structure. Neurites were searched by calculating the gradient vector flow[61] throughout the image. The neurites found in the search were then traced using a three-dimensional Sobel filter.[62] (**d**) The computer-generated model was manually examined and edited according to the method used in protein crystallography. The obtained working model was refined with conjugate gradient minimization. The geometric parameters were calculated from the three-dimensional Cartesian coordinates of the refined model. (**e**) Stereo drawing of a spiny dendrite indicated with the arrow head in panel **d**. The structure is superposed on a three-dimensional map of the observed image. The map drawn in gray is contoured at 2.5 times the standard deviation (2.5 σ) of the image intensity with a grid size of 96.6 nm.

Neurites in three-dimensional tissue can be regarded as three-dimensional curves. A three-dimensional curve can be represented with two parameters: curvature and torsion. Curvature corresponds to the reciprocal of the radius of the curve. Torsion represents the deviation of the curve from a plane. The neurites were divided into segments at each ramification point. The geometries of these neurite segments were analyzed by evaluating their curvature and torsion (Table 1). Spine structures were analyzed using the parameters of length, minimum radius, and maximum radius in addition to the curvature and torsion (Supplementary Table S4), since dendritic spines have been classified into several categories in terms of their neck width and length.[42,43] Spine density was defined as the number of spines per total length of spiny dendrites (Supplementary Tables S1 and S3).

**Neurite structures**

The curvatures and torsions of the neurite segments are summarized in Table 1. A total of 2737 neurite segments from the schizophrenia cases and 2254 segments from the control cases were analyzed. Figures 2a and 2b show the distributions of the curvature and torsion of the segments. The curvature distribution of the schizophrenia cases exhibited long upper tails (Fig. 2a), showing a 45% increase on average. This resulted in a larger standard deviation in the curvature of the neurite segments of the schizophrenia cases (Table 1; 0.28–0.36 µm$^{-1}$) in comparison with that of the control cases (0.21–0.23 µm$^{-1}$). The curvature median exhibited a significant difference even between the four schizophrenia cases ($p < 2.2 \times 10^{-16}$ with the Kruskal-Wallis test) and between the four control cases ($p < 2.2 \times 10^{-16}$). In contrast, the torsion showed no apparent difference between the schizophrenia and control cases (Fig. 2b). There was no significant difference in torsion median between all the cases ($p = 0.44$ with the Kruskal-Wallis test). The torsion distribution of every case has a peak at the origin, indicating that the neurites have no chiral bias (Fig. 2b).

**Table 1.** Geometric parameters of neurites.

| Case | Curvature (µm$^{-1}$) | | Torsion (µm$^{-1}$) |
|---|---|---|---|
| | Total | Orphan neurite | |
| S1 | 0.46 (0.28) / 523 | 0.58 (0.30) / 288 | -0.03 (0.35) / 513 |
| S2 | 0.47 (0.32) / 754 | 0.59 (0.34) / 450 | -0.02 (0.37) / 742 |
| S3 | 0.60 (0.34) / 435 | 0.78 (0.32) / 238 | 0.01 (0.41) / 426 |
| S4 | 0.71 (0.36) / 880 | 0.79 (0.36) / 700 | 0.00 (0.33) / 873 |
| N1 | 0.33 (0.22) / 415 | 0.38 (0.21) / 154 | 0.02 (0.44) / 389 |
| N2 | 0.44 (0.21) / 731 | 0.49 (0.21) / 422 | 0.01 (0.32) / 721 |
| N3 | 0.37 (0.21) / 491 | 0.46 (0.21) / 289 | 0.01 (0.27) / 484 |
| N4 | 0.41 (0.23) / 432 | 0.48 (0.24) / 252 | -0.03 (0.34) / 426 |

Values represent mean (sample standard deviation) / number of observations. S1–S4 are the schizophrenia cases, and N1–N4 are the control cases. The orphan neurite column represents statistics of neurites whose somata were outside of the viewing field.

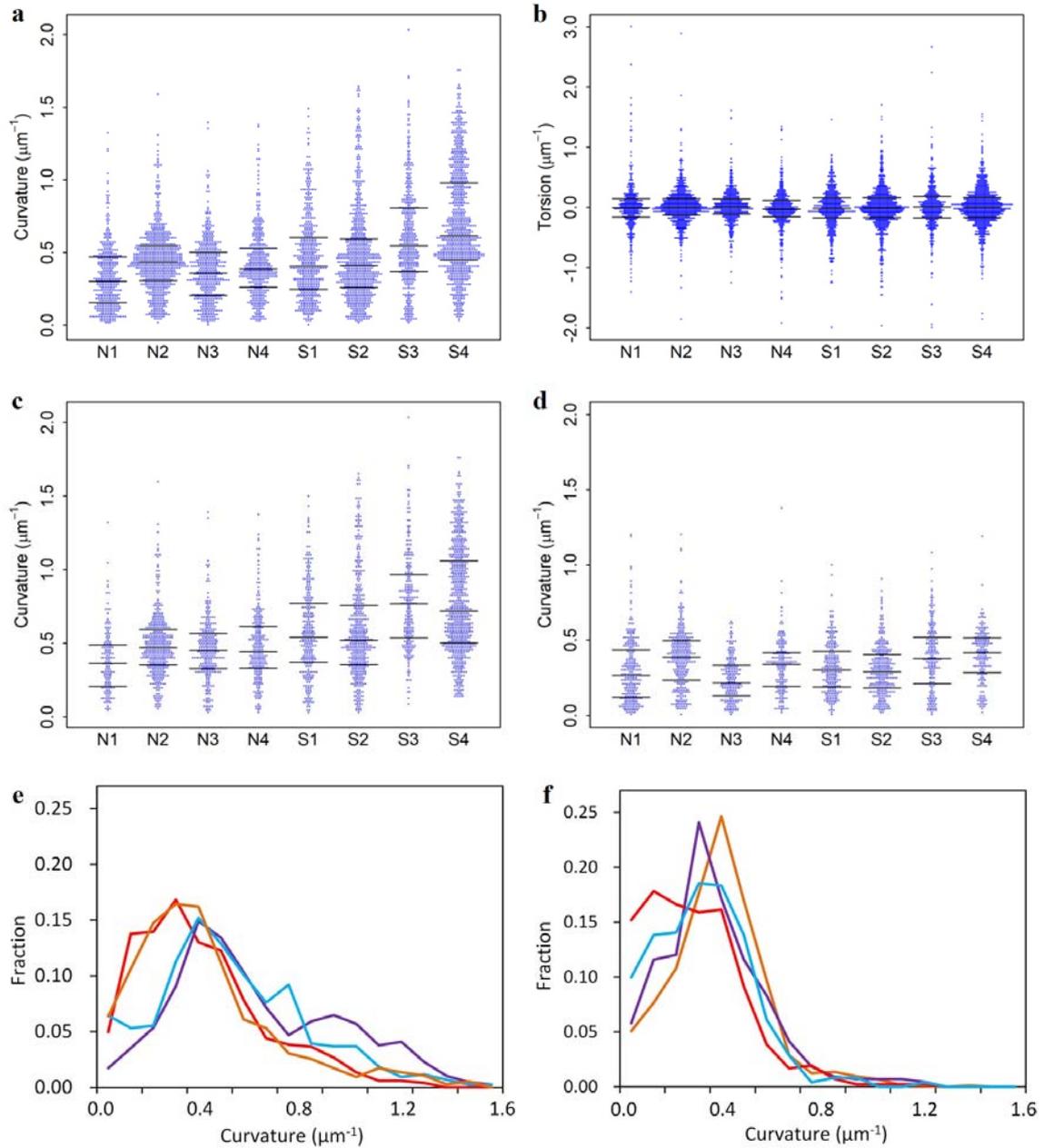

**Figure 2.** (**a**) Distribution of neurite curvature. Quartiles are indicated with bars. (**b**) Distribution of neurite torsion. (**c**) Distribution of curvature of orphan neurites without soma in the viewing field. (**d**) Distribution of curvature of neurites whose somata were visualized within the image. (**e**) Relative frequency of neurite in each 0.1 μm$^{-1}$ bin of curvature. The schizophrenia S1 case is plotted in red, S2 in orange, S3 in cyan, and S4 in purple. (**f**) Neurite curvature of the control N1 case is plotted in red, N2 in orange, N3 in cyan, and N4 in purple.

The neurites whose somata were visualized in the image are proximal ones within a viewing field width of 64–122 μm (Supplementary Table S2). Besides those laterally proximal structures, neurites without soma were also visualized. These orphan neurites should be distal parts of neuronal arbors whose somata were out of the viewing field. These two categories of neurites are separately plotted in Figs. 2c and 2d. The orphan neurites of the schizophrenia cases showed a wide curvature distribution, a 51% increase on average compared with the controls (Table 1). The mean curvatures of the orphan neurites were significantly different between the schizophrenia and control cases (Welch's t-test, $p = 0.020$, 4 schizophrenia and 4 controls). These results indicate that distal neurites have high curvature in the case of schizophrenia.

The profiles shown in Figures 2a–d vary between cases, even within the control cases, indicating that neuronal structures vary between individuals. The curvature median showed significant differences between the four control cases and also between the four schizophrenia cases, as described above. Figure 2e–f plots the relative frequency of the neurite curvatures. The frequency profiles were not identical between cases (Fig. 2e and 2f). The structures of the S2, S3, N1, and N3 cases were analyzed using multiple samples. The relative frequencies of the curvature in these multiple samples are separately plotted in Supplementary Figure S5. Multiple samples from the same individual have similar profiles, except that two samples of the N1 case show differences probably due to its long post-mortem time. These results suggest that the neuronal structures have features characteristic of each individual. A difference in neurite curvature results in a difference in the spatial trajectory and hence alters the neuronal circuits. The tissues analyzed in this study were taken from the anterior cingulate cortex. It has been shown that the anterior cingulate cortex has emotional and cognitive functions.[30,31] Therefore, the structural differences of the neurons of the anterior cingulate cortex should represent mental individuality or the psychiatric states of individuals.

Figures 3a and 3b, along with Supplementary Figure S6, show scatter plots of the mean curvature of the neurite trajectory and the mean radius of the neurite. The plots indicate reciprocal relationships between the trajectory curvature and the neurite radius. The schizophrenia cases show a wide distribution (Fig. 3a) for both spiny dendrites and smooth neurites, while the control cases show a narrow distribution (Fig. 3b). These plots indicate that the high-curvature neurites of the schizophrenia cases have a short radius.

Figure 3c shows representative structures of the neurites of the schizophrenia and control cases. The neurites of the schizophrenia cases exhibit frequent changes in direction (Fig. 3c), resulting in tortuous structures. The S4 neurite is thinner than the other neurites. In contrast, the neurites of the control cases show gradual and broad curves. The tissue structures of the S4 case and its age/gender-matched control N4 case are shown in Fig. 3d and 3e, respectively (also in Supplementary Videos S1 and S2). The structure of the S4 case is frizzy, whereas that of the N4 case is mostly straight. The S4 patient showed severe schizophrenic symptoms and bore a frame shift mutation in the GLO1 gene.[44,45] It has been shown that the GLO1 mutation can cause oxidative stress.[46] Therefore, the structural alteration of the neurites of the S4 case is ascribable to oxidative stress due to the GLO1 mutation.

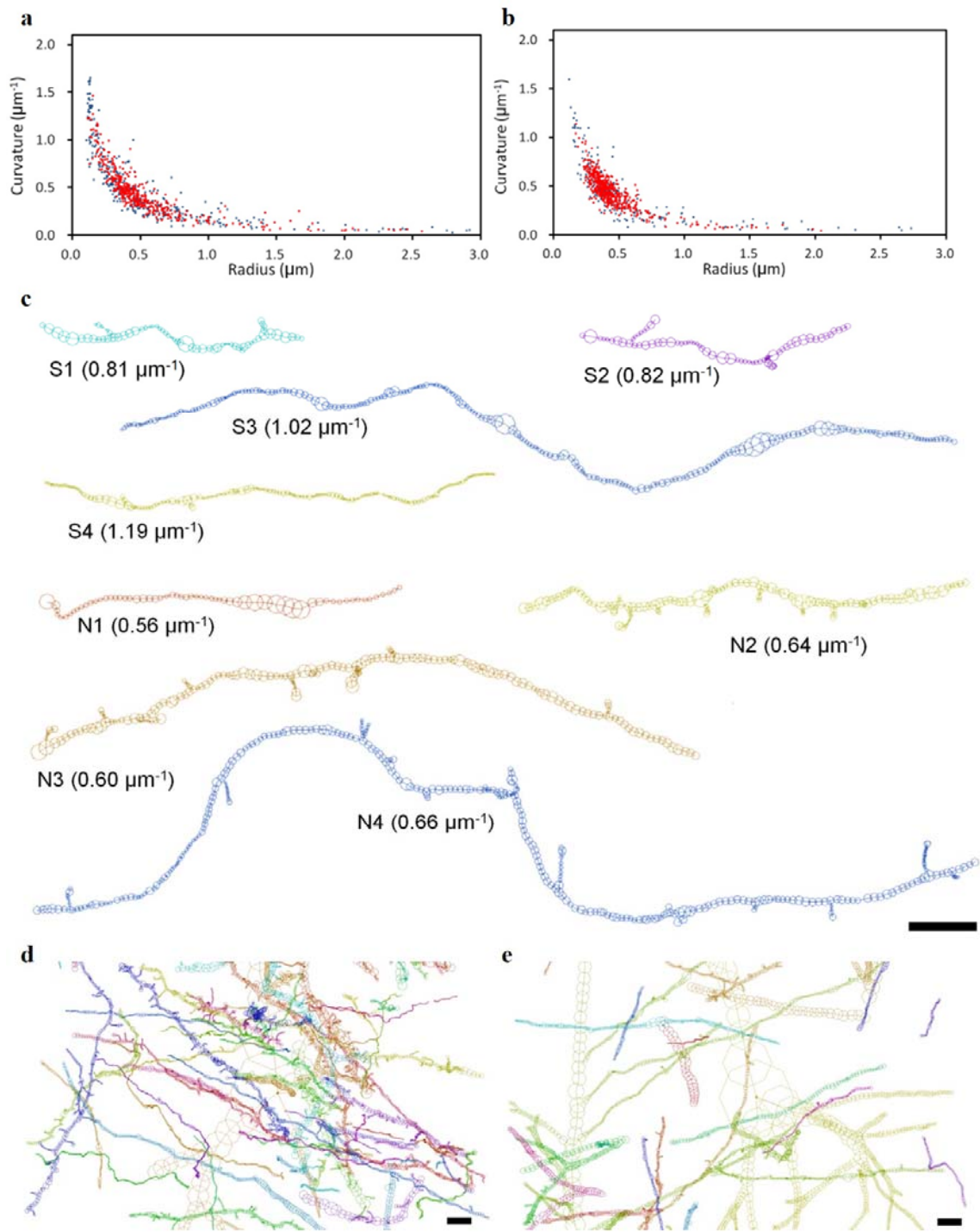

**Figure 3.** Neurites in schizophrenia and control cases. (**a**) Scatter plot of curvature and radius of neurites in the schizophrenia S2 case. (**b**) Scatter plot of control N2 case. The horizontal axis represents the mean radius of the neurite as a fiber. Thin neurites are on the left, and thick ones are on the right. The vertical axis represents the mean curvature of the neurite trajectory. Spiny dendrites are indicated with red dots and smooth neurites with blue. Neurites of which mean radii are larger than 3 μm are omitted. (**c**) Neurite

segments showing median curvature in the top quartile of each case. Mean curvature of each neurite is shown in parenthesis. The neurite of N1 is a branch on an apical dendrite of a pyramidal neuron. Others are orphan neurites of which somata are not visualized within the image. (**d**) Schizophrenia S4A structure. (**e**) Control N4A structure. Panels **d** and **e** were produced by placing the soma node of the largest pyramidal neuron at the figure center. The pial surface is toward the top. Structures are color-coded. Scale bars: 5 μm.

**Dendritic spine structures**

The geometric analyses of the dendritic spines are summarized in Supplementary Table S4. A total of 15116 spines of schizophrenia cases and 12885 spines of control cases were analyzed. It has been reported that dendritic spines can be categorized into several groups, such as mushroom spine and stubby spine.[42] Thin-necked and long spines can be characterized with a small minimum radius and long length, while stubby spines can be characterized with a large minimum radius and short length. Scatter plots of the spine length and minimum radius are shown in Fig. 4 and Supplementary Fig. S7. The plots of the schizophrenia and control cases have similar profiles composed of a wedge-shape domain and a triangle domain (Fig 4a, 4c, Supplementary Fig. S7). This indicates that dendritic spines can be categorized into two groups. Spines in the wedge domain exhibit short lengths and hence correspond to stubby spines. Their radius exhibits a nearly linear correlation to the length. Spines of the triangle domain exhibit long lengths and comparably short minimum radii and hence correspond to necked mushroom spines. The triangle distribution indicates that a longer spine has a thinner neck (Fig. 4a and 4c). The presence of a neck affects the ratio between maximum and minimum radii. However, the scatter plots of the radius ratio and spine length (Fig. 4b, 4d, Supplementary Fig. S7) show no distinctive domains. This indicates that the neck morphology is continuous.[43] The schizophrenia and control cases show similar profiles in all the plots, suggesting that structures of dendritic spines in the schizophrenia cases analyzed in this study were the same as those of the controls.

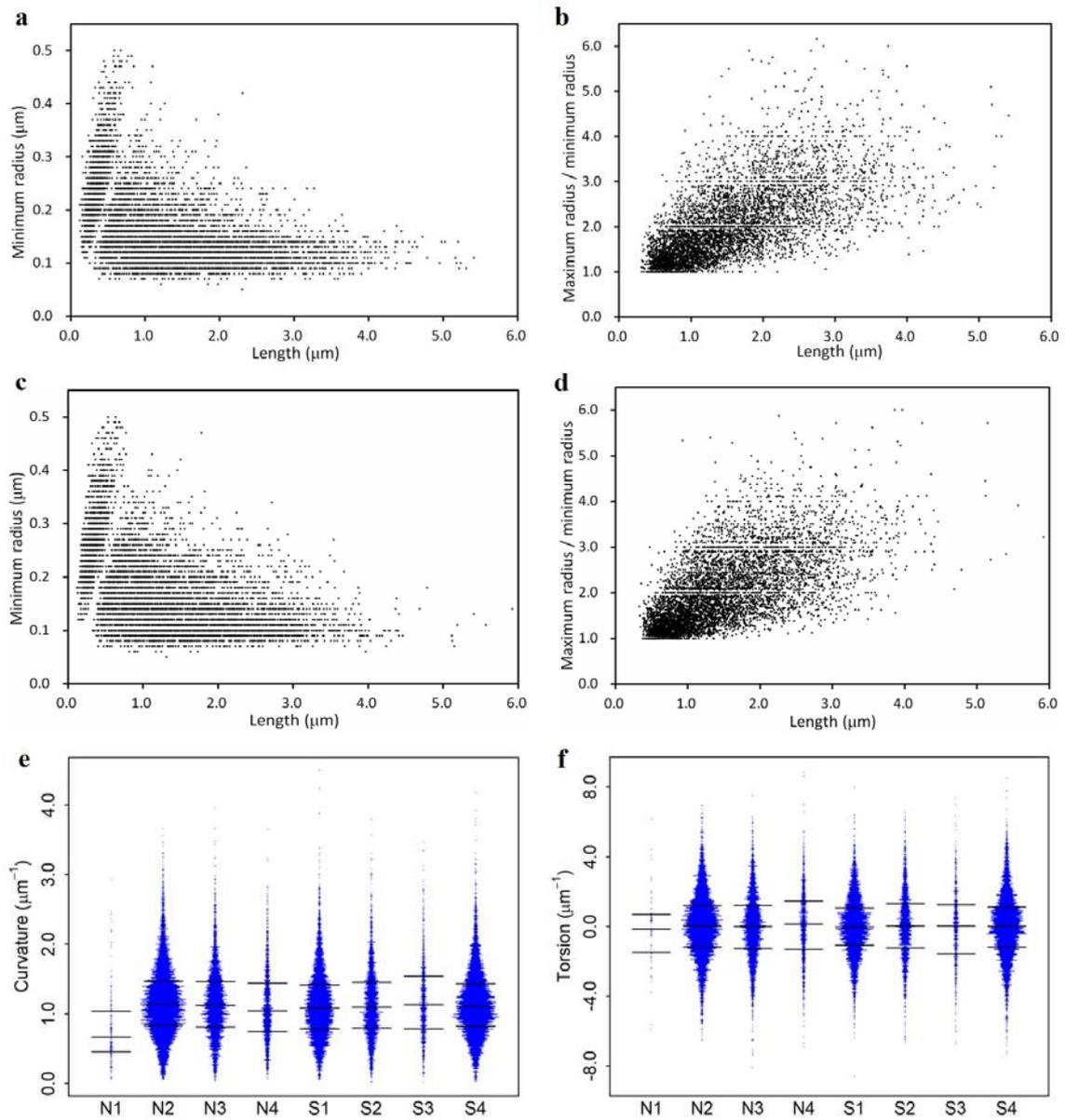

**Figure 4.** (**a**) Scatter plot of spine parameters in the schizophrenia S4 case. Minimum node radius is plotted against length. Three outliers (length/radius = 7.51/0.13, 6.24/0.10, and 6.23/0.10) were omitted. (**b**) Radius ratio between maximum / minimum radii of S4 is plotted against length. Four outliers (length/ratio = 7.51/2.77, 6.24/2.90, 6.23/3.70, and 3.55/7.43) were omitted. (**c**) Minimum radius of N2. An outlier (length/radius = 7.88/0.14) was omitted. (**d**) Radius ratio of N2. An outlier (length/ratio = 7.88/3.14) was omitted. (**e**) Distribution of spine curvature. Quartiles are indicated with bars. (**f**) Distribution of spine torsion.

It has been reported that spine density in the external pyramidal layer (layer III) of the frontal cortex[2] and the temporal cortex[3] is significantly lower in schizophrenia cases than in control cases. It has also been reported that the difference in spine density is not significant in the internal pyramidal layer of the frontal cortex[47] or in the external pyramidal layer of the occipital cortex.[2] In this study, we analyzed spine structures mainly in the internal pyramidal layer of the anterior cingulate cortex. The obtained spine density per dendritic length is summarized in Supplementary Table S1. The spine density was comparable to those observed in the Golgi-stained frontal cortex.[48] No significant difference in mean spine density was found between the four schizophrenia and four control cases (Welch's t-test, $p = 0.56$), though the sample size of this study is small. The spine density of the N1 case was lower than in the other cases, presumably due to the long post-mortem time. Except for this case, the curvatures and torsions of the spines showed similar distribution profiles (Fig. 4e and 4f), suggesting that the schizophrenia and control cases analyzed in this study share common spine structures.

**Discussion**

In the S4 schizophrenia case, the high curvature and short radius of the neurites should stem from the GLO1 frameshift mutation.[45] Although similar structural alterations were also observed in the S1, S2, and S3 schizophrenia cases, their causes are not clear at present. Some adverse effects similar to oxidative stress of the GLO1 mutation should have degraded the neuronal structures. Tortuous neurites have been observed in the cerebral cortex of schizophrenic brain.[49] It has been reported that the schizophrenia-susceptible DISC1 protein[50] interacts with a number of factors associated with neuronal functions.[51] Apical dendrites of dentate gyrus neurons showed morphological alterations in mice carrying a DISC1 mutation.[52] A significant decrease in dendritic diameter was reported for a Shn2 knockout mouse with schizophrenia-like symptoms.[53] The disruption of any susceptible genes related to the neuronal structure or environmental risk factors that affect brain development can alter geometry of neurites.

*N*-methyl-D-aspartate (NMDA) receptor antagonists including phencyclidine cause psychiatric symptoms similar to those of schizophrenia.[54] A corkscrew deformity of dendrites was reported in an animal model of the NMDA receptor hypofunction.[55] These reports suggest that the high-curvature neurites observed in this study are related to schizophrenia symptoms. However, we cannot exclude the possibility that antipsychotics affected the neuronal structures. Such drug effects can be elucidated by using nanotomography to analyze brain tissues of drug-treated animals.

The neurite of the schizophrenia cases showed higher curvature and shorter radius compared with the controls. The radius of the neurite affects its conductivity, resulting in altered connections between neurons. This neurite thinning should have a relationship to the tissue volume reduction observed in schizophrenia.[6-8] It has been suggested that a neurodevelopmental defect in the neuropil can explain the loss of cortical volume without loss of neurons.[56] The curvature of a neurite determines its spatial trajectory. A curve with a higher curvature reaches more positions in its three-dimensional vicinity, but needs to be longer to reach a distal position compared with a straight line. This can alter the connectivity of the neuronal circuit. It would be difficult to relieve or restore these nanometer-scale structural alterations of the tissue. Therefore, the deteriorative outcome of structurally altered neurons should be prevented in advance of their incorporation into the neuronal circuit. The results obtained in this study hence support the consensus that early diagnosis and treatment is important for a better prognosis of schizophrenia.

A disadvantage of x-ray visualization of brain tissue is that neurons show little contrast in x-ray images, since they are composed of light elements. In this study, cerebral tissues were stained with Golgi impregnation in order to label their neurons with silver. Therefore, the obtained results are tempered by the limitations of the staining method. Since only a small number of neurons are stochastically visualized in Golgi impregnation, the labelled neurons are limited representatives of the neuronal population.[57] The viewing field width and number of cases are also limitations. Although hundreds of neurites and thousands of spine structures were analyzed for each case, the results reported in this paper are of millimeter-sized tissues of the anterior cingulate cortex of four schizophrenia and four control cases. Another limitation of our case-control study is that the controls were not psychiatrically evaluated. There is a possibility that the control cases had latent mental diseases, although no geometric hallmark of schizophrenia was observed in them.

Biological individuality is encoded in the genome. At the macroscopic level, individuals are identified from the body structure, such as face, fingerprints, or overall brain connectivity.[58] However, at the cellular level, little evidence of mental personality has been delineated on the basis of neuronal structures, though histological studies suggested differences of cellular structures between individuals.[59,60] The results reported in this paper reveal differences in the tissue structure of the anterior cingulate cortex between schizophrenia and control cases, and also between control individuals. This suggests that geometric profiles of brain tissue are different between individuals. Structural differences of cerebral neurons can result in differences in brain circuits, and hence affect the mental individuality.

Human mental activities are performed through coordination of many diverse areas of the brain and cannot be explained from one study of a single area. The temporal lobe has been reported to show a tissue volume reduction in schizophrenia[9], and hence, it should be analyzed with the same strategy. The results of this study also suggest that humans have nanometer- to micrometer-scale structural diversity in the cerebral cortex. Although the present sample size was sufficient for evaluating the statistical significance of the structural difference, schizophrenia is a complex and heterogeneous psychiatric disorder.[1,9] Therefore, the differences observed in this study should be re-examined by analyzing more three-dimensional structures of cerebral tissues of schizophrenia and control cases. Such further analyses will lead to a better understanding of our mental individuality and to better diagnosis and treatment of schizophrenia.

**Conflict of interest**
The authors declare no conflict of interest.

**Acknowledgements**
We are grateful to Prof. Motoki Osawa and Akio Tsuboi (Tokai University School of Medicine) for their generous support of this study. We are also grateful to Prof. Yasuo Ohashi (Chuo University; Statcom Co., Ltd.) for his helpful advice regarding the statistical tests. We appreciate Prof. Yoshiro Yamamoto (Tokai University) for his helpful advice regarding the data analysis. We also appreciate Dr. Jun Horiuchi (Tokyo Metropolitan Institute of Medical Science) for his suggestions regarding the manuscript. We thank Noboru Kawabe (Support Center for Medical Research and Education, Tokai University) for assistance in preparing the histology sections. We also thank the Technical Service Coordination Office of Tokai University for assistance in preparing adapters for nanotomography. This work was supported by Grants-in-Aid for Scientific Research from the Japan Society for the Promotion of Science (nos. 21611009, 25282250, and 25610126). The synchrotron radiation experiments at SPring-8 were performed with the approval of the Japan Synchrotron Radiation Research Institute (JASRI) given to R.M. (proposal nos. 2011B0034, 2012B0034, 2013A0034, 2013B0034, 2013B0041, 2014A1057, 2015A1160, 2015B1101, 2017A1143, 2018A1164, and 2018B1187). The synchrotron radiation experiments at the Advanced Photon Source of Argonne National Laboratory were performed during the 2016-2 and 2017-3 runs under General User Proposal GUP-45781 by R.M. This research used resources of the Advanced Photon Source, a U.S. Department of Energy (DOE) Office of Science User Facility operated for the DOE Office of Science by Argonne National